\documentclass[%
 aip,
 amsmath,amssymb,
 reprint,%
]{revtex4-1}

\usepackage{graphicx}
\usepackage{dcolumn}
\usepackage{bm}
\usepackage{amssymb}   
\usepackage{amsmath, bm}
\usepackage{tikz,tikz-3dplot} 

\usepackage[utf8]{inputenc}
\usepackage[T1]{fontenc}
\usepackage{etoolbox}

\newcommand{\abs}[1]{ \left\lvert#1\right\rvert}

\makeatletter
\def\@email#1#2{%
 \endgroup
 \patchcmd{\titleblock@produce}
  {\frontmatter@RRAPformat}
  {\frontmatter@RRAPformat{\produce@RRAP{*#1\href{mailto:#2}{#2}}}\frontmatter@RRAPformat}
  {}{}
}%
\makeatother
\begin{document}

\preprint{AIP/123-QED}

\title{Pure circularly polarized light emission from waveguide microring resonators}
\author{Leonardo Massai}
\author{Tom Schatteburg}%
\author{Jonathan P. Home}
\author{Karan K. Mehta$^*$}
\altaffiliation{School of Electrical and Computer Engineering, Cornell University, Ithaca, NY 14853, USA}
\affiliation{$^1$Department of Physics \& Institute for Quantum Electronics, ETH Z{\"u}rich, 8093 Z\"urich, Switzerland}%

\email{karanmehta@cornell.edu.}

\date{\today}

\begin{abstract}
Circularly polarized light plays a key role in many applications including spectroscopy, microscopy, and control of atomic systems. Particularly in the latter, high polarization purity is often required. Integrated technologies for atomic control are progressing rapidly, but while integrated photonics can generate fields with pure linear polarization, integrated generation of highly pure circular polarization states has not been addressed. Here, we show that waveguide microring resonators, perturbed with azimuthal gratings and thereby emitting beams carrying optical orbital angular momentum, can generate radiated fields of high circular polarization purity. We achieve this in a passive device by taking advantage of symmetries of the structure and radiated modes, and directly utilizing both transverse and longitudinal field components of the guided modes. On the axis of emission and at maximum intensity, we measure an average polarization impurity of $1.0 \times 10^{-3}$ in relative intensity across the resonance FWHM, and observe impurities below $10^{-4}$ in this range. This constitutes a significant improvement over the ${\sim}10^{-2}$ impurity demonstrated in previous work on integrated devices. Photonic structures allowing high circular polarization purity may assist in realizing high-fidelity control and measurement in atomic quantum systems.
\end{abstract}

\maketitle

Many applications of integrated photonics currently under research rely on launching guided modes in waveguide structures to tailored free-space propagating beams. These include optical phased arrays \cite{sun2013large, komljenovic2017sparse, wang20192d}, systems for neural stimulation and control \cite{hoffman2016low, mohanty2020reconfigurable}, and control of atomic quantum systems \cite{mehta2016integrated, mehta2020integrated, niffenegger2020integrated, kim2018photonic, newman2019architecture, spektor2022universal}, among others. Many functions in atomic systems in particular require high circular polarization purity (${\gtrsim}30$ dB polarization extinction ratio), for example selective optical pumping into particular states via polarization-selective excitation \cite{haffner2008quantum}, driving of closed cycling transitions for qubit state readout \cite{langer2006high, kwon2017parallel}, or certain implementations of laser cooling \cite{roos2000experimental, lechner2016electromagnetically}. Pure polarization states are often key in precision measurement experiments \cite{steffen2013note} and polarization-sensitive spectroscopy as well \cite{10.1103/physrevlett.36.1170, 10.1364/ao.56.000b92}. In integrated photonic implementations, simple grating devices can by symmetry naturally couple single quasi-TE or TM waveguide modes to free-space beams with areas of highly pure linear polarization \cite{mehta2017precise}; but generation of high-purity circular polarization has so far not been addressed.

Previous work towards circular polarization emitters has included two-dimensional gratings fed from two perpendicular inputs with equal amplitude and active phase control to appropriately superpose two linear polarizations \cite{10.1364/ol.39.002553}. Due to fabrication imperfections, such approaches require active phase tuning, and the challenge of balancing amplitude and phase simultaneously limited purity to ${\sim}99\%$. These challenges would likewise affect realizations based on conventional grating couplers simultaneously emitting both quasi-TE and TM modes of a single feed waveguide, which in principle offer another route to circularly polarized emission. The circular symmetry of integrated structures emitting beams carrying orbital angular momentum (OAM) \cite{doerr2011circular, 10.1109/jstqe.2019.2941488} may assist in realizing pure circular polarization, but studies of polarization impurities from such structures have been limited to the few-percent level \cite{cogne2020generation}. 

\begin{figure*}[t!]
\centerline{\includegraphics[width=.9\textwidth]{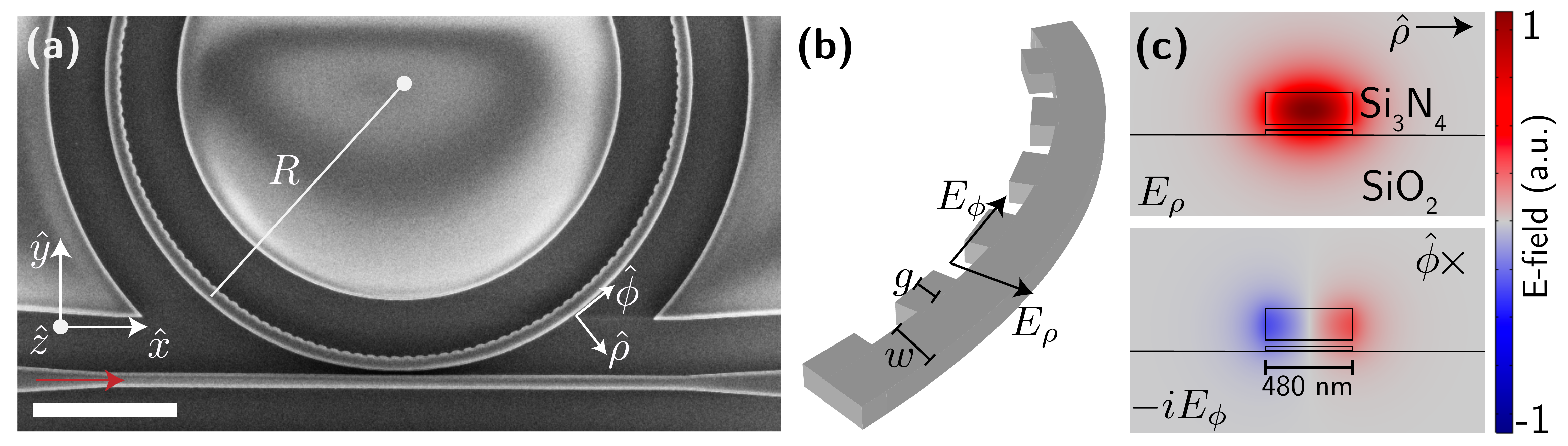}}
\vspace{0 cm}
\caption{\label{fig:schem} (a) SEM image of the fabricated device. Scale bar: 5 $\mu$m.  (b) Schematic of section of the resonator, showing waveguide width and grating amplitudes $w$ and $g$, and directions of relevant mode field components.  (c) $\bm{\hat \rho}$ and $\bm{\hat \phi}$ field components of the fundamental mode guided in the ring with designed dimensions $w + g = 480$ nm and $g=30$ nm. The waveguides are clad in SiO$_2$, on a Si substrate (not shown) 2.7 $\mu$m below the bottom of the Si$_3$N$_4$ \cite{mehta2019towards}. }
\end{figure*}

Our work demonstrates passive waveguide microresonator devices in which the symmetry of the device and emitted modes, together with the vector nature of the guided modes, ensures high purity of circularly polarized emission at the intensity maximum of the emitted beam. Our device (shown in Fig.~\ref{fig:schem}) is based on a ring resonator perturbed by an azimuthal grating, similar to structures which were shown in previous work to emit beams carrying OAM \cite{cai2012integrated}. We briefly describe the principle for emission of pure circular polarization on axis, and proceed to the experimental observations on fabricated devices. 

We consider a ring with $n$ periods of a grating introduced around its circumference; a resonant mode of index $m$ corresponds to a guided electric field $\bm{E_g} \propto e^{i m \phi}$ which we initially assume to be polarized primarily along the radial direction $\bm{\hat \rho}$. The structure is perturbed by the grating to emit into radiating modes with field $\bm{E_r}$. With $\Delta \epsilon$ representing the perturbation, the coupling coefficient can be written as $\kappa \propto \int dV \bm{E_r}^* \Delta \epsilon \bm{E_g}$. \cite{liu2009photonic} Considering a perturbation with $n$ periods around the circumference and taking its lowest harmonic, $\Delta \epsilon \propto \cos(n\phi)$. Expanding $\bm{E_r}$ in a basis of radially polarized modes with azimuthal phase dependence $e^{i \eta \phi}$, we see that $\bm{E_r}$ has nonzero contributions from modes with $\eta  = m\pm n$. Since $m+n$ will for practical values be a large number corresponding to high-OAM beams that diffract strongly, the emission near axis will be heavily dominated by light with $e^{il\phi}$ phase dependence, with $l \equiv m-n$. We can decompose, for example, the radial component of the radiated field $\propto \bm{\hat \rho} e^{il\phi}$, in the circular polarization basis defined by $\bm{\hat \sigma_\pm} \equiv (\bm{\hat x} \mp i\bm{\hat y})/\sqrt 2$. In the paraxial limit, the eigenmodes in this basis would correspond to the Laguerre-Gauss modes \cite{allen1992orbital}. Since $ \bm{\hat \rho}  \cdot \bm{\hat \sigma}_\pm = e^{\mp i\phi}/\sqrt 2$, we see that radially polarized radiation with azimuthal phase factor $e^{il\phi}$ can be written as $\propto \bm{\hat \sigma}_+ e^{i(l-1)\phi} + \bm{\hat \sigma}_- e^{i(l+1)\phi}$. In this basis, owing to the phase singularity on axis associated with modes with azimuthally varying phase, only modes with 0 azimuthal phase factor have nonzero intensity on axis. Thus, for $l=1$, only $\bm{\hat \sigma}_+$, or pure right-hand circular polarization (RHCP) is radiated on axis. 

In addition to this on-axis nulling of one circular polarization component due to the structure of the emitted OAM modes, in our device the vector nature of the guided modes of the ring serves to further enhance the total power emitted into a desired circular polarization, even off-axis. The fundamental quasi-TE waveguide mode in the ring (Fig.~\ref{fig:schem}c) consists not only of the dominant $\bm{\hat \rho}$-polarized electric field, but also an azimuthal component along $\bm{\hat \phi}$, with an amplitude that grows in proportion to the lateral confinement. While weaker than the radial component for our configuration, the azimuthal field's magnitude is maximal near the waveguide edges where the grating is introduced, and hence can be designed to contribute nearly equally to emission. Furthermore, from Gauss' law applied to a lossless mode field, the azimuthal component can be seen to oscillate with a $\pm \pi/2$ phase difference with respect to the dominant radial component, with $\pm$ corresponding to opposite sides of the guide. As a result, the displacement currents associated with the grating perturbations are naturally circularly polarized with respect to the $\bm{\hat z}$ direction, to a degree depending on the relative amplitude of the two field components at the grating location. This intrinsic circular polarization of the radiating dipoles is related to mechanisms used in previous work \cite{cogne2020generation}, though in the present work results primarily from the vectorial nature of the guided mode itself rather than design of emitting elements. As quantitatively described in Appendix A, this results in preferential emission of radiation in one circular polarization over the other. 

\begin{figure}[b]
\centerline{\includegraphics[width=.5\textwidth]{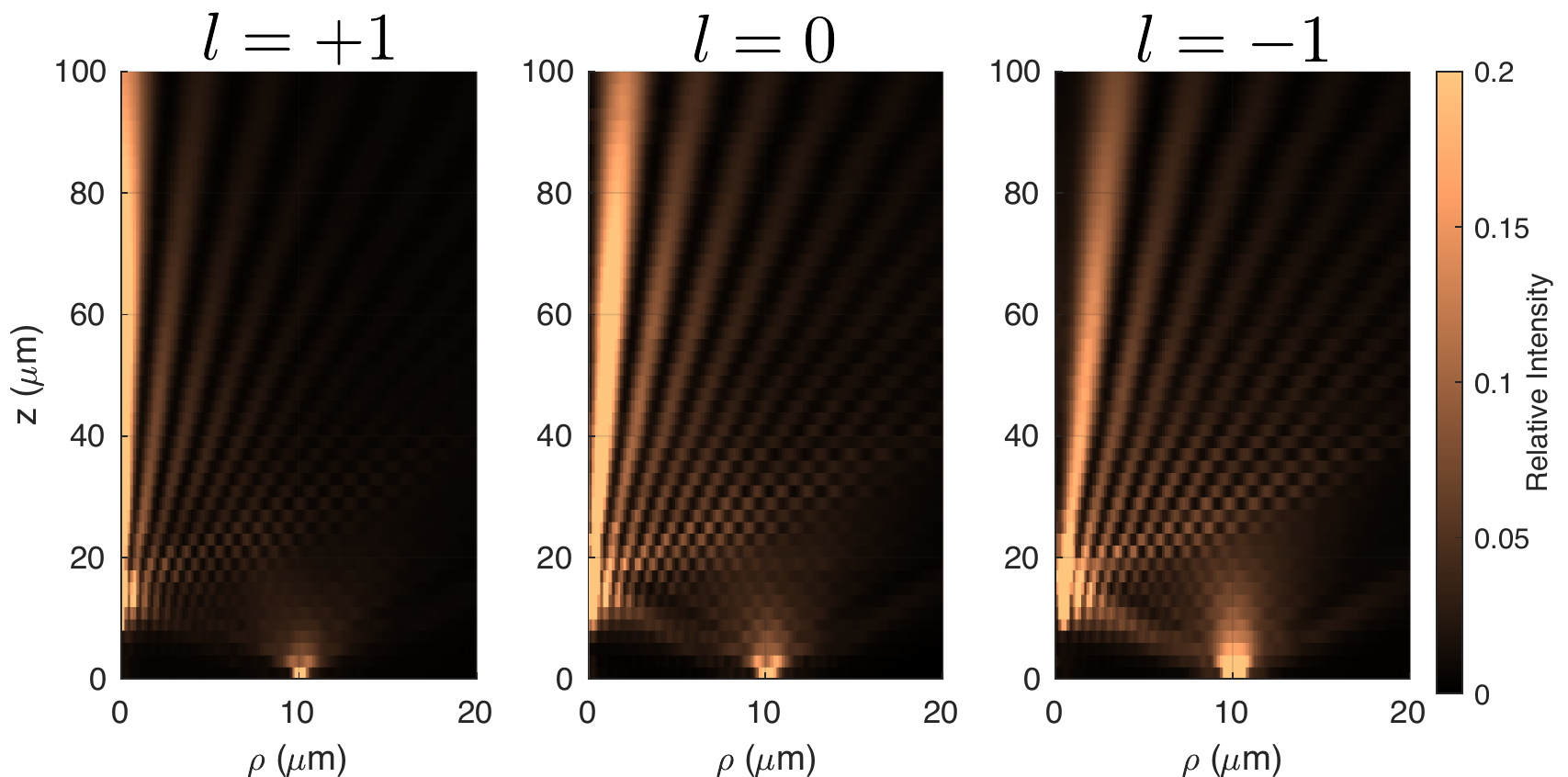}}
\vspace{0 cm}
\caption{\label{fig:heightscans} Measured cross-sections of total radiated intensity in the $\rho-z$ plane for the resonant modes, from left to right, with $l=1$ ($\lambda = 722.8$ nm); $l=0$ ($\lambda = 726.8$ nm); and $l=-1$ ($\lambda = 730.9$ nm). }
\end{figure}

Devices simultaneously leveraging both of the above effects were fabricated in a commercial foundry process \cite{worhoff2015triplex} on the same wafer and using the same layers as devices presented previously \cite{mehta2019towards, mehta2020integrated}. Devices tested were fabricated alongside components designed to deliver light addressing the optical qubit transition in $^{40}\mathrm{Ca}^+$ at $\lambda=729$ nm, and our ring devices were designed to function near this wavelength. 

Light is coupled to devices through single-mode fibers at edge couplers $\pm 2.7$ mm from the waveguide/ring coupling region, and sourced by a tunable Ti:Sapphire laser (M Squared SolsTiS) whose frequency is measured using a wavelength meter (HighFinesse WS-6). Devices are fixed to a temperature-controlled stage held at $25.0$ $^\circ$C. For a first set of measurements characterizing total emission regardless of polarization, radiated intensity is imaged through a 0.65 NA NIR objective (Mitutoyo 50$\times$ Plan Apo NIR HR) onto a scientific CCD sensor. 

\begin{figure*}[t!]
\centerline{\includegraphics[width=.9\textwidth]{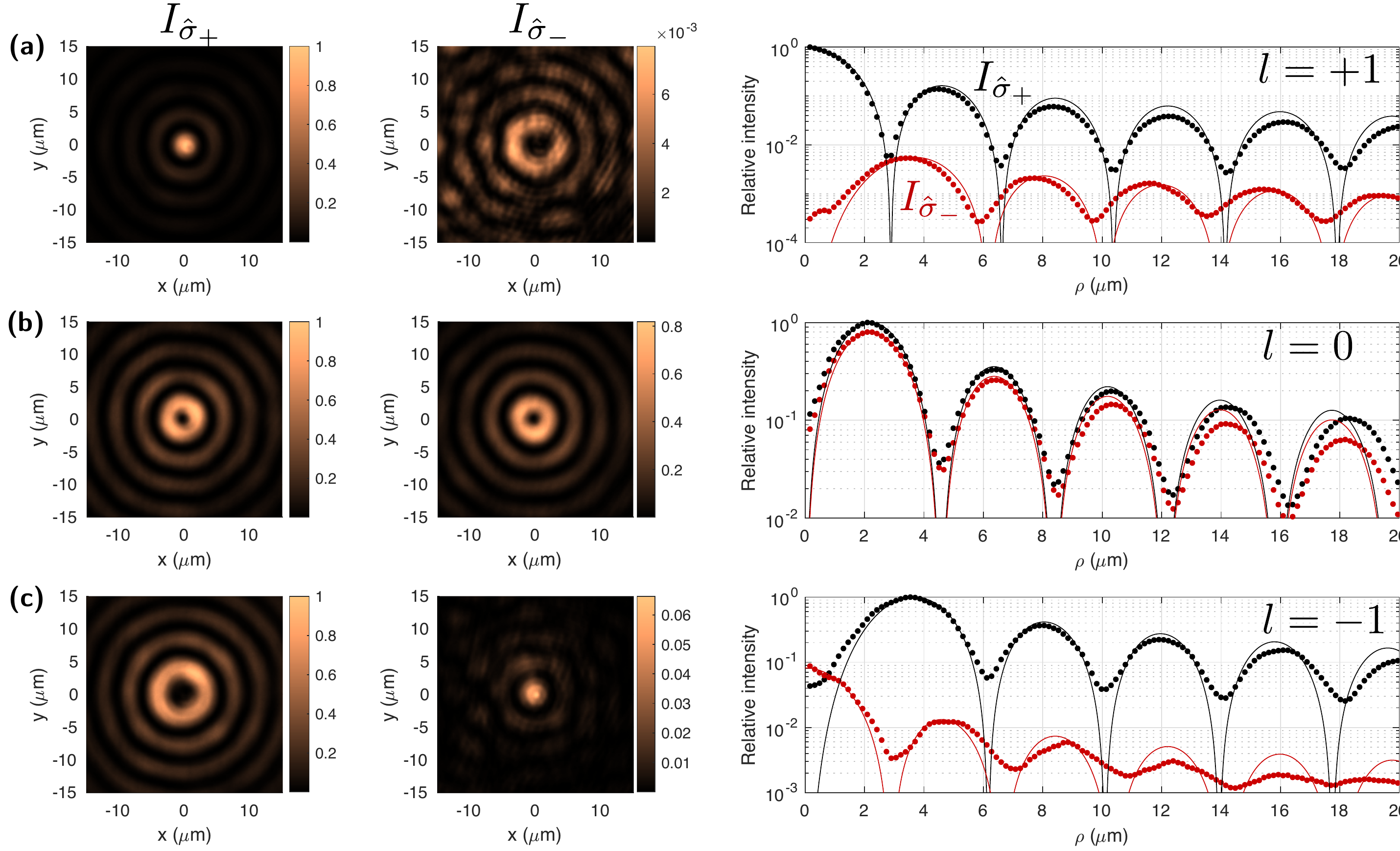}}
\vspace{0 cm}
\caption{\label{fig:polcomps} Measured intensity in both the $\bm{\hat \sigma_+}$ and $\bm{\hat \sigma_-}$ polarization components, imaged at $z = 100$ $\mu$m above the emitter for the (a) $l=+1$, (b) $l=0$, and (c) $l=-1$ modes. Points in the rightmost plots show the measured radial profiles, averaged around the azimuth, together with fits (lines) from the simple analytical diffraction model described in the text. }
\end{figure*}

The measured total intensity emitted is shown in Fig.~\ref{fig:heightscans}; for three modes of the ring with $l = 1$, $0$, and $-1$, a stack of images of intensity in the $x-y$ plane is obtained at various heights $z$, from which the intensity is averaged over the azimuth to yield the profiles in the $\rho-z$ plane. A clear maximum on-axis is seen for the $l=+1$ mode; the $l=0$ mode shows a null in intensity on axis due to the nonzero OAM of each circular polarization component emitted at this mode; and $l=-1$ shows nonzero but low intensity on-axis, due to the suppression of the $\bm{\hat \sigma_-}$ emission due to the guided mode structure.

To directly probe the polarization structure of the emitted radiation, intensity in either circular polarization component is imaged by inserting into the imaging setup a quarter-wave plate (QWP) and linear polarizer (LP), with the LP mounted in a motorized rotation stage. By appropriate tuning of the LP angle $\theta$, we select either $\bm{\hat \sigma_+}$ or $\bm{\hat \sigma_-}$ light to be transmitted to the image sensor. Fig.~\ref{fig:polcomps} shows the recorded intensities in both polarization components at $z=100$ $\mu$m above the ring structure, for the three resonant modes. We comment on the qualitative features of these plots, and then describe a simple analytical model below that describes this behavior. Fig.~\ref{fig:polcomps}(a) indicates the high purity on axis achieved for the $l=+1$ mode, while (c) shows similar mode profiles except with nonzero intensity on-axis now present in the $\bm{\hat \sigma_-}$ component, as expected based on the structure of the OAM modes described above. However, its power relative to the $\bm{\hat \sigma_+}$ component is significantly suppressed, owing to the enhancement of total power in the $\bm{\sigma_+}$ component resulting from the guided mode vector structure. The $l=0$ mode is uniquely characterized by a strong contradirectional coupling within the resonator, since second-order diffraction for this resonance is phase matched to reflection within the ring \cite{cai2012integrated}. As a result, this mode has a significantly larger amplitude of the resonant mode propagating in the $-\bm{\hat \phi}$ direction of Fig.\ref{fig:schem}(b) than the $l = \pm 1$ modes. Counterpropagating light in a given resonant mode can be seen to emit the opposite circular polarization as compared to the forward-propagating mode, and we hence observe nearly equal intensities in both polarization components for this mode. 

\begin{figure*}[t]
\centerline{\includegraphics[width=.8\textwidth]{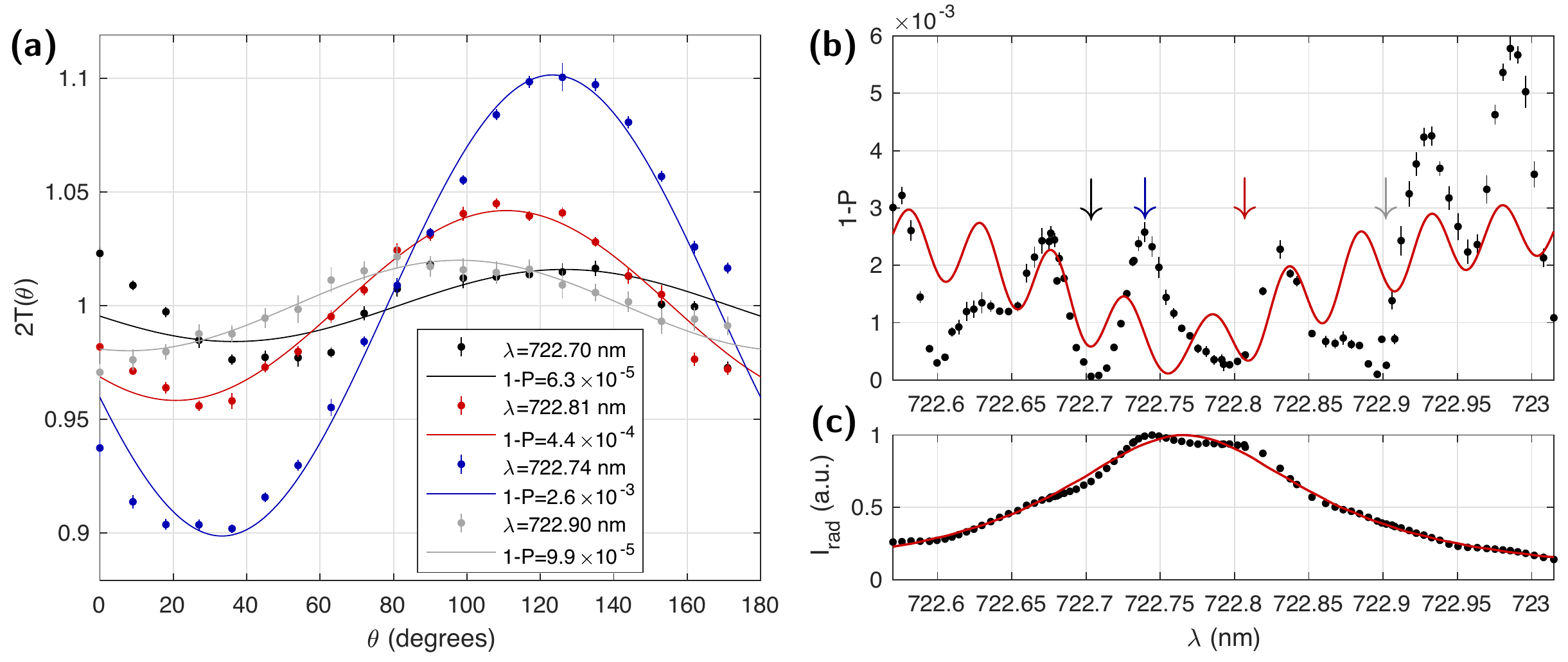}}
\vspace{0 cm}
\caption{\label{fig:purities} (a) Transmitted intensities on axis vs. rotation angle $\theta$ of linear polarizer, normalized to the average transmission of each scan, for four wavelengths within the FWHM. Fit contrast of these oscillations allows extraction of circular polarization impurity $1-P$, as labeled in the legend. Error bars on data points represent standard deviations of 20 measurements. (b) Fitted $1-P$ values across the $l=+1$ resonance, with the datasets shown in (a) labeled with the colored arrows. Error bars correspond to standard errors of the fitted purity. (c) Radiated intensity vs. wavelength. Red lines in (b) and (c) correspond to the output of the same simple model described in the Appendix which captures some qualitative features of the purity spectrum.}
\end{figure*}

These features are captured by a simple analytical diffraction model for the emitted intensity components. We extend a previously described model \cite{zhu2013theoretical} to the case of superposed azimuthally and radially polarized emitters. For the purposes of calculating the diffracted field, grating teeth are approximated as azimuthally distributed point dipoles, with complex amplitudes of radial (azimuthal) dipole moments proportional to the resonant mode field components $E_\rho$ ($E_\phi$). As shown in Appendix A, this results in an approximate radiated field for either circular polarization component: 

\begin{equation}
\begin{split}
\mathcal{E}_{ \hat{\sigma}_{\pm}, l}(\rho, \phi, z)=K_{\pm} \Phi(\rho, z) e^{i(l \mp 1) \phi}J_{l\mp1}(\zeta)\\\times\left(E_{\rho} \mp i E_{\phi}\right)
\end{split}
\label{emission model}
\end{equation}
where $J_m$ represents the $m$-th order Bessel function of first kind. The amplitude $K_\pm$ and phase factor $\Phi(\rho,z)$ are defined in the Appendix. 

The solid lines in the rightmost plots of Fig. 3 show the radial intensity profiles predicted by this model at the measurement height. The model for the $l = \pm 1$ modes are fit to the data with one free parameter (neglecting normalization), namely the ratio $\abs{E_\rho / E_\phi}$ describing the relative amplitude of the components of the guided mode at the grating teeth, which is found to be 1.34 (and common to all of the fits). The $l=0$ mode is fit with an additional parameter $A_r$ corresponding to the amplitude of the counter-propagating wave (which is absent for the other modes in this model). Despite the simplicity of this model, the general agreement with the measurement, particularly near the axis where the paraxial approximation employed is reliable, confirms our understanding of the mechanism for emission and the polarization structure of the radiated fields. 

We now proceed to quantify polarization purity on axis and at the point of maximum intensity for the $l=+1$ mode of interest. We measure circular polarization purity by observing intensity transmitted through a linear polarizer (LP) as a function of its angle. Consider an input propagating along $\bm{\hat z}$ with transverse electric field polarization described by the unit vector $\bm{\hat{p}} = \sqrt{P} \bm{\hat{\sigma}}_+ + e^{i\delta} \sqrt{1-P} \bm{\hat{\sigma}}_-$, the real $0\le P\le1$ representing circular polarization purity (here chosen with respect to the right-hand circularly polarized (RHCP) component $\bm{\hat \sigma}_+$) and $\delta$ a relative phase between the RHCP and LHCP components. The power transmission through a LP oriented to transmit electric field polarized along an angle $\theta$ with respect to $\hat x$ can be evaluated as 
\begin{equation} 
T(\theta) = \frac12\left(1 + 2\sqrt{P(1-P)} \cos(\theta - 2\delta) \right). 
\end{equation}
$2T(\theta)$ thus gives this transimssion normalized to its mean, and the contrast $C(P) \equiv 2\sqrt{P(1-P)}$ constitutes a highly sensitive measure of $P$. 

Fig.~\ref{fig:purities}(a) shows these oscillations in total intensity transmitted through a ${\sim} 1$ mm diameter pinhole aligned to the emission axis and positioned approximately 20 cm above the ring device, as a function of the linear polarizer angle $\theta$. Figs.~\ref{fig:purities}(b) and (c) show the emitted purity together with total emitted power across the resonance. Averaging the impurity across the resonance FWHM, we obtain a low average impurity of $1.0\times10^{-3}$. However, we clearly observe significant dependence of the purity on the wavelength within the resonance. We attribute this dependence to undesired reflections in the structure, resulting in light in the ring counterpropagating to that of the intended resonance and thereby emitting the opposite circular polarization. A few observations indicate counterpropagating light limited the measured purity: (1) In device variants with larger coupling gaps between the bus and ring, and as a result with higher total $Q$, consistently lower purities were measured. Standard models of contra-directional coupling in rings \cite{little1997surface} indicate that for the same reflection strength within the ring (in our case, e.g. from the grating teeth themselves), the fraction of reverse-to-forward propagating power scales strongly with increasing $Q$. These observations are consistent with counterpropagating light contributing higher impurities in higher-$Q$ devices. (2) We also observed that devices with larger grating amplitudes exhibited lower purities; this is again consistent with impurity due to back-reflection within the ring arising from the grating teeth themselves. (3) Some qualitative features of the purity spectrum could be captured by a simple model accounting for intra-ring as well as facet reflection in our test chips; this model, described in Appendix B, resulted in the approximate fit lines in Fig.~\ref{fig:purities}(b) and (c). These considerations motivate minimizing grating-induced back-reflection, e.g. with a sinusoidal modulation of the waveguide width instead of the rectangular grating elements used in the present devices, and simultaneously minimizing back-reflections from any facets or interfaces in the structures used to couple to the device. 

Full FDTD solutions of the structure tested indicate that 26\% of the power guided in the ring is radiated in the $+\bm{z}$ direction. The measured transmission within the bus at the $l = +1$ resonance is 37\%, indicating that up to 63\% of the input power is coupled into the ring. The total efficiency of the current device with respect to upwards radiated power is therefore estimated to be approximately 17\%. We note that the thickness of the SiO$_2$ between the waveguides and substrate in the present devices was not optimized for maximal upwards radiation, and could be to ensure constructive interference of the direct emission along $\bm z$ with the substrate reflection \cite{mehta2017precise}. 

In addition to minimizing undesired reflections and thereby impurity, future designs may address other limitations of the present devices: namely, only radiation near the axis possesses maximal purity, emission is constrained to be normal to the chip surface, and significant power is carried in sidelobes. The ratio between $|E_\rho|$ and $|E_\phi|$ was not deliberately optimized in this work; optimizing waveguide and perturbation geometry to set this more closely to unity would allow a higher suppression of undesired polarization, across the extent of the beam profile. An array of such optimized emitters from e.g. parallel straight waveguides, each with gratings on one side, would no longer benefit from the OAM mode-structure associated purity of the present devices, but may bring valuable compensating features, including the possibility for high purity emission at flexible angles even without a resonant structure, and beam-forming to minimize undesired sidelobes via tuning of the emitters' relative exictation amplitudes. This would furthermore alleviate the need for precise tuning of the device to the desired operating wavelength, a drawback of resonant approaches. 

This work demonstrates high circular polarization purity emission from integrated optical structures. The ideas presented here may play a role in structures interfacing to atomic systems where such purity can be a critical requirement. Future work will explore devices utilizing emission from both transverse and longitudinal field components of the guided modes to attain similar purities in broadband nonresonant devices, as well as across larger areas of the emitted beam. Integration of such devices into ion trap or neutral atom systems promises a richer set of integrated functionalities as compared to those achievable with linear polarizations, and may advance realizations of the varied possibilities for tailored atom-light interactions in structured lightfields \cite{schmiegelow2012light, solyanik2019excitation, vasquez2021ion}. 

\begin{acknowledgments}
We thank LioniX Intl. for fabrication of the devices, and Fabian Kaufmann for lending the NIR objectives used for imaging measurements. KKM acknowledges an ETH postdoctoral fellowship that supported the early phase of this work. 
\end{acknowledgments}

\section*{Data Availability Statement}
The data that support the findings of this study are available from the corresponding author upon reasonable request.

\subsection*{Author contributions}
T.S. performed simulations of the structure, and specified designs. L.M. characterized fabricated devices experimentally, developed the analytical radiation model, and performed data analysis. K.K.M. conceived the work, drew designs and coordinated fabrication, and supervised the project. K.K.M. and L.M. wrote the manuscript with input from all authors. 

\appendix
\section{Emission model}
\begin{figure}[t]
\centering
\includegraphics[scale=0.25]{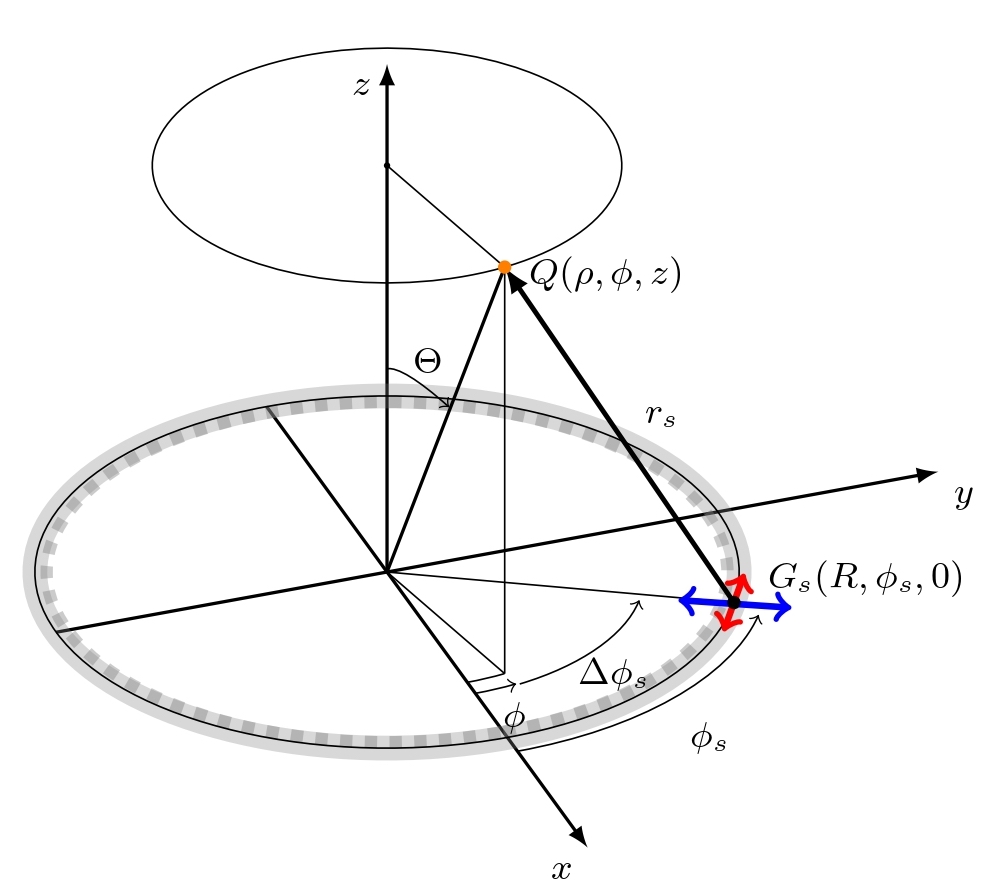}
\caption{Schematic of the calculation setup, to find the field at any point $Q$. The ring has a radius $R$, $n$ grating teeth around the circumference at positions $G_s$, each of which is modeled as a point dipole source with azimuthal and radial components. }
\label{fig:scheme}
\end{figure}

Here we develop an analytical diffraction model to describe the radiated electric field $\bm{\mathcal{E}}(\rho,\phi,z)$ at any point in space $(\rho,\phi,z)$ in cylindrical coordinates with basis vectors $\hat{\bm{\rho}}=\cos\phi \hat{\bm{x}}+\sin\phi\hat{\bm{y}}, \hat{\bm{\phi}}=-\sin\phi\hat{\bm{x}}+\cos\phi\hat{\bm{y}}, \hat{\bm{z}}$. This model extends the one developed in previous work \cite{zhu2013theoretical} to our case of emitting dipoles with moments along both the radial and azimuthal directions. All coordinates and the parameters used are indicated in Fig.~\ref{fig:scheme}.  

We describe radiation from a resonant mode at wavelength $\lambda $ and with free-space wavenumber $k_0=2\pi/\lambda$, with a field in the ring proportional to $e^{i m \phi}$ as in the main text. The grating teeth lie at points $G_s=(R,\phi_s,0)$ with $\phi_s=2\pi s/n$ and $s=1,2,...,n$. Treating these teeth as perturbations to the resonant mode, we  model the emitted electric field as sourced by effective point dipoles with moments $\bm{P}_s=( P_{\phi}\hat{\bm{\phi}}_s+ P_{\rho}\hat{\bm{\rho}}_s)e^{i\phi_s l}$, situated at the grating teeth locations $G_s$, and with $l=m-n$ as in the main text. The complex components of the effective radiating dipole moments $P_\phi$ and $P_\rho$ are driven by the corresponding components $E_{\phi}$ and $E_{\rho}$ of the unperturbed ring waveguide mode-field, according to $\bm{P} = V \Delta \epsilon \bm{E}$. Here, $\Delta \epsilon$ is the difference between the core and cladding permittivity (and therefore the amplitude of the perturbation), and $V$ is a scaling factor with units of volume accounting for the spatial extent of the perturbation.  

The vectorial expression of the radiated electric field can be written as superposition of the dipole fields sourced by $P_{\phi}$ and $P_{\rho}$ in the far-field limit (dropping terms $\propto1/r^2$ and $1/r^3$):
\small
\begin{equation}
    \begin{split}
    \bm{\mathcal{E}}_{l}(\rho, \phi, z)=&\mathcal{E}_{\hat{\rho},l}\hat{\bm{\rho}}+\mathcal{E}_{\hat{\phi},l}\hat{\bm{\phi}}+\mathcal{E}_{\hat{z},l}\hat{\bm{z}}\\=&A\sum_{s=1}^{n}e^{ik_0 r_s}e^{i\phi
    _s l}\frac{k_0^2}{r_s}(\hat{\bm{r}}_s \times \hat{\bm{\phi}}_s)\times \hat{\bm{r}}_s\\
    +&B\sum_{s=1}^{n}e^{ik_0 r_s}e^{i\phi_s l}\frac{k_0^2}{r_s}(\hat{\bm{r}}_s \times \hat{\bm{\rho}}_s)\times \hat{\bm{r}}_s\\
    \end{split}
    \label{eq:E_init}
\end{equation}
\normalsize
where $A=P_{\phi}/4\pi\epsilon_0= V\Delta\epsilon_r E_{\phi}/4\pi$ and $B=P_{\rho}/4\pi\epsilon_0=V\Delta\epsilon_r E_{\rho}/4\pi$. 

Considering the far-field zone with $\rho/z \equiv \tan\Theta \ll 1$, we can approximate
\begin{equation*}
    \begin{split}
         r_s=& \sqrt{z^2+(R^2+\rho^2-2\rho R\cos\Delta\phi_s)}\\
         \simeq&z+\frac{(R^2+\rho^2)}{2z}-R\tan\Theta\cos\Delta\phi_s
    \end{split}
    \end{equation*}
where $\Delta\phi_s=\phi_s-\phi$, as shown in Fig. \ref{fig:scheme}.\\
Defining the global phase $\Phi(\rho,z) \equiv e^{ik_0 \left(z+\frac{R^2+\rho^2}{2z}\right)}$ we can write:
    \begin{equation}
        e^{i k_0 r_s}=\Phi(\rho,z)e^{-ik_0 R\tan\Theta \cos\Delta\phi_s}.
    \end{equation}
and we approximate $1/r_s \approx 1/z$.

We can then express the transverse field components of Eq. \ref{eq:E_init} within these approximations as
\small
\begin{equation}
\begin{split}
&\mathcal{E}_{\hat{\rho},l}(\rho, \phi, z)=\Phi(\rho, z)\frac{k_0^{2}}{z} \\ &\times  
\sum_{s=1}^{n} e^{-ik_0 R\tan\Theta \cos \Delta\phi_s} e^{i l\phi_s} (-A\sin \Delta \phi_s+B \cos \Delta \phi_s )\\
&\mathcal{E}_{\hat\phi, l}(\rho, \phi, z)=\Phi(\rho, z)\frac{k_0^{2}}{z} \\ &\times \sum_{s=1}^{n} e^{-ik_0 R\tan\Theta \cos \Delta\phi_s} e^{i l\phi_s} (A\cos \Delta \phi_s+ B\sin \Delta \phi_s)\\
\end{split}
\end{equation}
\normalsize
and the longitudinal component as
\small
\begin{equation}
\begin{split}
&\mathcal{E}_{\hat z, l}(\rho, \phi, z)=\Phi(\rho, z)\frac{k_0^{2}}{z^2} \\ &\times 
\sum_{s=1}^{n} e^{-ik_0 R\tan\Theta \cos \Delta\phi_s} e^{i l\phi_s} \big(A\rho\sin \Delta \phi_s+  B\left(R-\rho\cos \Delta \phi_s\right)\big).
\end{split}
\end{equation}
\normalsize

From now on, we will consider only the transverse field components $\mathcal{E}_{\hat\rho, l}$ and $\mathcal{E}_{\hat\phi, l}$, since $\mathcal{E}_{\hat z, l}$ decays as $\sim1/z^2$. 
With a basis change, we can find the $\hat{\bm{\sigma}}_+$ and $\hat{\bm{\sigma}}_-$ components of the transverse field according to 
\begin{align}
\left(\begin{array}{c}
  \mathcal{E}_{\hat{\sigma}_+,l} \\
  \mathcal{E}_{\hat{\sigma}_-,l}  \\
\end{array}\right) 
= \frac{1}{\sqrt{2}}\left[\begin{array}{cc}
 e^{-i\phi}  & -ie^{-i\phi}  \\
 e^{i\phi}  & ie^{i\phi}   \\
\end{array}\right]
\left(\begin{array}{c}
  \mathcal{E}_{\hat\rho,l}  \\
  \mathcal{E}_{\hat\phi,l}  \\
\end{array}\right), 
\end{align}
resulting in: 
\begin{equation}
\begin{split}
\mathcal{E}_{\hat{\sigma}_+, l}(\rho, \phi, z)&=\Phi(\rho, z)\frac{k_0^{2}}{z\sqrt{2}} e^{-i \phi} (B-iA)\\ &\times  
\sum_{s=1}^{n} e^{-ik_0 R\tan\Theta \cos \Delta\phi_s} e^{i l\phi_s} e^{-i\Delta \phi_s} \\
\mathcal{E}_{\hat{\sigma}_-, l}(\rho, \phi, z)&=\Phi(\rho, z)\frac{k_0^{2}}{z\sqrt{2}} e^{i \phi} (B+iA)\\ &\times  \sum_{s=1}^{n} e^{-ik_0 R\tan\Theta \cos \Delta\phi_s} e^{i l\phi_s} e^{i\Delta \phi_s}.
\end{split}
\end{equation}

By multiplying and dividing $e^{il\phi}$ from $\mathcal{E}_{\hat{\sigma}_\pm, l}$, we can use an approximate integral representation of the Bessel function of the first kind (since $2\pi/n \ll 1$):\\
\begin{equation}
\sum_{s=1}^{n} e^{\left(i \zeta \cos \Delta \phi_s\right)} e^{ \left(i h\Delta \phi_s\right)} \approx i^{h} n J_{h}(\zeta)
\end{equation}
\\
where $\zeta \equiv -k_0 R\tan \Theta=-k_0 R\rho/z$.
We arrive at a more compact form for the two transverse field components:
\begin{equation}
\begin{split}
\mathcal{E}_{\hat{\sigma}_\pm, l}(\rho, \phi, z)&=\Phi(\rho, z)\frac{nk_0^{2}}{z\sqrt{2}} e^{i(l\mp1) \phi} \\ &\times  
i^{l\mp1} J_{l\mp1}(\zeta) (B\mp iA)\\
\end{split}
\end{equation}

Substituting $A$ and $B$ to express the radiated field in terms of the guided electric field components $E_{\phi}$ and $E_{\rho}$, we obtain the final expression for the radiated field, used in Eq. \ref{emission model}:\\
\begin{equation}
\begin{split}
\mathcal{E}_{ \hat{\sigma}_{\pm}, l}(\rho, \phi, z)=K_{\pm} \Phi(\rho, z) e^{i(l \mp 1) \phi}J_{l\mp1}(\zeta)\\\times\left(E_{\rho} \mp i E_{\phi}\right)
\end{split}
\label{eq:model}
\end{equation}
with $K_{\pm}=\frac{i^{l\mp1} k_0^2 n V \Delta\epsilon_r}{4 \sqrt{2} \pi z}$. We see that with $E_{\rho}/E_\phi = \pm i$, the $\bm{\sigma_\pm}$ emission component is nulled and power is radiated only in $\bm{\sigma_\mp}$. 

Though in our devices the two field components were not designed to be perfectly matched in amplitude ($|E_\rho/E_\phi| = 1.34$ according to the fits in Fig.~\ref{fig:polcomps}), the resulting power suppression ratio $\left| \frac{E_\rho + i E_\phi}{E_\rho - i E_\phi} \right|^2$ evaluates to approximately 0.02, indicating the robustness of this  suppression even when $|E_\rho| = |E_\phi| $ is not perfectly satisfied. 

\section{Multi-reflection model}

To qualitatively understand the wavelength-dependence of the purity observed in Fig.~\ref{fig:purities}(b), a simple model accounting for multiple reflections within our device was analyzed and fit to the data. Since counterpropagating light in the ring emits into the undesired circular polarization, the model's aim is to calculate the counterpropagating field amplitude. The impurity then expected from the counterpropagating field is given by $\abs{E_-}^2/(\abs{E_+}^2 + \abs{E_-}^2)$, where $E_+$ ($E_-$) is the forward (counter-) propagating field amplitude in the ring. 

\begin{figure}[t]
\centerline{\includegraphics[width=.5\textwidth]{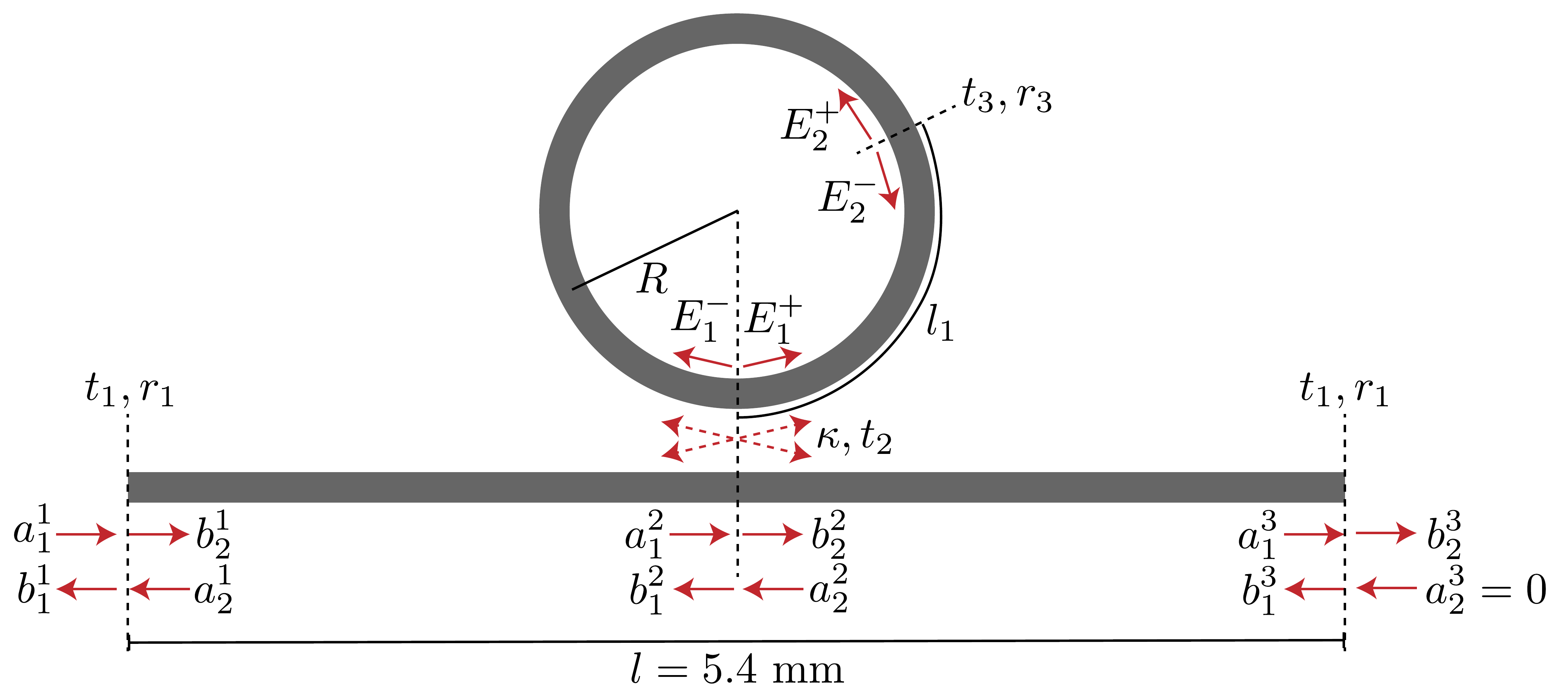}}
\vspace{0 cm}
\caption{\label{fig:rmodel} Schematic of field amplitudes solved for in the multiple-reflection model. The free parameters are $r_1$, $\kappa$, $r_3$, and $f$.}
\end{figure}

The model incorporates reflections from the facets of the device, separated by $l=5.4$ mm on the die and with reflection coefficient $r_1$; reflection within the ring with real amplitude $r_3$ at an effective position $2\pi R f$ along the ring ($f$ satisfies $0<f<1$ and effectively sets the phase of the intra-ring reflection); and coupling between bus and ring with coefficient $\kappa$; as drawn in Fig.~\ref{fig:rmodel}. The transmission coefficients are $t_{1,3} = \sqrt{1-r_{1,3}^2}$, $t_2 = \sqrt{1-\kappa^2}$.  Our model therefore has four free real parameters: $r_1$, $r_3$, $f$, and $\kappa$. 

From the periodicity of the grating and the measured resonance wavelength of the $l = 0$ mode, the measured ring effective index is $n_\mathrm{eff} = 1.657$ at $\lambda=726.8$ nm (within 1\% of that simulated for the target waveguide geometry) and from the measured FSR the group index is $n_\mathrm{g} = 2.078$; these parameters are fixed in the fit. 

The various mode amplitudes are solved for according to the following system of equations, with reference to Fig.~\ref{fig:rmodel}: 
\begin{eqnarray*}
b_2^1 &=& -i t_1 a_1^1 - r_1 a_2^1 \\
b_1^2 &=& t_2 a_2^2 - i \kappa E_2^- e^{i\theta_1 - \alpha l_1} \\ 
b_2^2 &=& t_2 a_1^2 - i \kappa E_2^+ e^{i\theta_2 -\alpha l_2} \\
b_1^3 &=& -r_1 a_1^3 \\ 
E_1^+ &=& -i\kappa a_1^2 + t_2 E_2^+ e^{i\theta_2 - \alpha l_2}  \\
E_1^- &=& -i\kappa a_2^2 + t_2 E_2^- e^{i\theta_1 - \alpha l_1} \\
E_2^+ &=& t_3 E_1^+ e^{i\theta_1 - \alpha l_1} -i r_3 E_1^- e^{i\theta_2 - \alpha l_2}  \\
E_2^- &=& t_3 E_1^- e^{i\theta_2 - \alpha l_2} -i r_3 E_1^+ e^{i\theta_1 - \alpha l_1}
\end{eqnarray*}
where $a_1^2 = b_2^1 e^{i \beta l/2}$ and so on for the other components propagating along the bus. $l_1 = 2\pi R f$ is the distance of the effective scatterer from the coupling region, and $l_2 = 2\pi R (1-f)$. $\alpha$ represents attenuation in the ring, including the loss due to the grating, and is set in the fit by matching the total ring $Q$ (including loss due to coupling $\kappa$) to that measured. $\theta_{1,2} = n_\mathrm{eff}k_0l_{1,2}$ are the phases associated with propagation around the two segments of the ring. 

The red lines in Fig.~\ref{fig:purities}(b) and (c) respectively correspond to the impurity and total stored power for the fit result with $r_1 = 0.063$, $r_3=0.0023$, $f = 0.97$, and $\kappa = 0.31$. Though the model is too simple to capture all features observed in our measured purity spectrum, it does approximately reproduce the periodicity of the oscillations, which are set by the measured effective indices, known device dimensions, and the interference of both the facet and intra-ring reflection (these oscillations are not present when either reflection is neglected). While a more complex model with more effective reflectors from possible defects e.g. in the bus would allow a more faithful fit, this relatively simple model suggests the importance of intra-ring reflection together with reflections back to the structure in limiting purity in the current device, and thereby routes to improvement in future implementations.

\bibliography{RingGrating}

\end{document}